\documentclass[twocolumn,showpacs,preprintnumbers,superscriptaddress,amsthm,amsmath,amssymb]{revtex4}
\usepackage{graphicx}
\usepackage{dcolumn}
\usepackage{bm}
\usepackage{color}
\usepackage{booktabs,longtable}
\usepackage{multirow}
\usepackage{CJK}
\usepackage{ltxtable}
\usepackage{rotating}
\usepackage[colorlinks,linkcolor=blue,anchorcolor=blue,citecolor=red]{hyperref}

\begin{document}
\begin{CJK*}{GBK}{song}

\title{Green's function method for the single-particle resonances in a deformed Dirac equation}

\author{T.-T. Sun}
\email{ttsunphy@zzu.edu.cn}
\affiliation{School of Physics, Zhengzhou University, Zhengzhou 450001, China}

\author{L. Qian}
\affiliation{School of Physics, Zhengzhou University, Zhengzhou 450001, China}

\author{C. Chen}
\affiliation{School of Physics, Zhengzhou University, Zhengzhou 450001, China}

\author{P. Ring}
\affiliation{Physik-Department der Technischen Universit\"{a}t M\"{u}nchen, D-85748 Garching, Germany}

\author{Z. P. Li}
\email{zpliphy@swu.edu.cn}
\affiliation{School of Physical Science and Technology, Southwest University, Chongqing 400715, China}

\date{\today}

\begin{abstract}
Single-particle resonances are crucial for exotic nuclei near and beyond the drip lines. Since the majority of nuclei are deformed, the interplay between deformation and orbital structure near threshold becomes very important and can lead to an improved description of exotic nuclei. In this work, the Green's function (GF) method is applied to solve the coupled-channel Dirac equation with quadrupole-deformed Woods-Saxon potentials for the first time. The detailed formalism for the partial-wave expansion of the Green's function is presented. A new approach getting exact values for energies and widths of resonant states by the GF method is proposed. Numerical checks are carried out by comparing with our previous implementation of the spherical GF method and the results from the deformed complex momentum representation~(CMR), the analytical continuation of the coupling constant (ACCC), and the scattering phase shift (SPS) methods, and it is proved that the GF method is very effective and reliably for describing resonance states, no matter they are narrow or broad, spherical or deformed. Finally, Nilsson levels for bound and resonant orbitals in the halo candidate nucleus $^{37}$Mg are calculated from the deformed GF method over a wide range of deformations and some decisive hints of $p$-wave halo formation are shown in this nucleus, namely, the crossing between the configurations $1/2[321]$ and $5/2[312]$ at deformation parameter $\beta>0.5$ may enhance the probability to occupy the $1/2[321]$ orbital that is originated from the $2p_{3/2}$ shell.
\\


\end{abstract}

\maketitle
\end{CJK*}

\section{\label{sec:intr}Introduction}

Since the experimental discovery of the neutron halo phenomenon in $^{11}$Li~\cite{PRL1985Tanihata_55_2676}, the structure of exotic nuclei, especially those close to the neutron or proton drip line, has attracted wide attentions in theory and experiment~\cite{PPNP1996PRing_37_193,PR2004Jonson_389_1,PR2005Vretenar_409_101,PPNP2006MengJ_57_470,PPNP2013Tanihata_68_215,JPG2015JMeng_42_093101}. In these drip-line nuclei, since the neutron or the proton Fermi surface is very close to the continuum threshold, the valence nucleons can be easily scattered by the pairing correlations to single-particle resonant states in the continuum and thus the coupling between the bound states and the continuum becomes very important~\cite{PRC1996Dobaczewski_53_2809,PRL1996Meng_77_3963,PRL1997Poschl_79_3841,PRC2015Changizi_91_024305,PRC2015Qi_92_051304}. Unexpected properties very different from those of normal nuclei have been observed or predicted, such as halo phenomenona~\cite{PRL1996Meng_77_3963,PRL1997Poschl_79_3841}, giant halos~\cite{PRL1998MengJ_80_460,PRC2002MengJ_65_041302, PRC2003Sandulescu_68_054323,SCP2003ZhangSQ_46_632,PRC2006TerasakiJ_74_054318,PRC2006Grasso_74_064317}, new magic numbers~\cite{PRL2000Ozawa_84_5493,PRL2018Michimasa}, and deformed halos~\cite{PRC2010Hamamoto_81_021304,PRC2010Zhou_82_011301,PRC2012Li_85_024312,PLB2014SSZhang_730_30,PLB2018XXSun_785_530}. Therefore, the resonant states close to the continuum threshold are essential for the investigation of exotic nuclei, and are also closely relevant to the nucleosynthesis of chemical elements in the universe~\cite{PRC2012SSZhang_86_032802,PRC2015Faestermann_92_052802}. Moreover, because the majority of nuclei far from stability are deformed, the interplay between deformation and near-threshold orbitals becomes very important as the shell structure evolves with deformation.

Based on conventional scattering theories, many approaches, such as R-matrix theory~\cite{PR1947Wigner_72_29,PRL1987Hale_59_763}, K-matrix theory~\cite{PRC1991Humblet_44_2530}, S-matrix theory~\cite{Book1972Taylor-ScatteringTheor,PRC2002CaoLG_66_024311}, the Jost function approach~\cite{PRL2012LuBN_109_072501, PRC2013LuBN_88_024323}, and the scattering phase shift (SPS) method~\cite{Book1972Taylor-ScatteringTheor,PRC2010LiZP_81_034311,SCP2010Li_53_773} have been developed to study the single-particle resonant states. Meanwhile, the techniques for bound states have been extended for the single-particle resonant states, such as the analytical continuation of the coupling constant~(ACCC) method~\cite{Book1989Kukulin-TheorResonance,PRC1997Tanaka_56_562,PRC1999Tanaka_59_1391,PRC2000Cattapan_61_067301,
CPL2001YangSC_18_196,PRC2004ZhangSS_70_034308,PRC2005Guo_72_054319,EPJA2007SSZhang_32_43,PRC2012SSZhang_86_032802,
EPJA2012SSZhang_48_40,EPJA2013SSZhang_49_77,PLB2014SSZhang_730_30,PRC2015Xu_92_024324}, the real stabilization method~(RSM)~\cite{PRA1970Hazi_1_1109,PRL1993Mandelshtam_70_1932,PRA1999Kruppa_59_3556,NPA2004Hagino_735_55,PRC2008ZhangL_77_014312}, the complex scaling method~(CSM)~\cite{PR1983Ho_99_1,PRC1986Gyarmati_34_95,PRC1988Kruppa_37_383,PRL1997Kruppa_79_2217,PRC2006Arai_74_064311,
CPC2010Guo_181_550,PRC2010JYGuo_82_034318,PRC2014ZLZhou_89_034307,PRC2015MShi_92_054313,
PRC2012QLiu_86_054312,PRC2014Shi_90_034319}, the complex-scaled Green's function (CGF) method~\cite{PLB1998Kruppa_431_237,PTP2005Suzuki_113_1273,PRC2016Shi_94_024302,EPJA2017Shi_53_40}, and the complex momentum representation~(CMR) method~\cite{NPA1968Berggren_109_265,PRC1978Kwon_18_932,JPA1979Sukumar_12_1715,PRC2006Hagen_73_034321,PRL2016Li_117_062502,PRC2017Fang_95_024311,PRC2017Tian_95_064329}. Especially, the SPS~\cite{PRC2010LiZP_81_034311,SCP2010Li_53_773}, ACCC~\cite{PRC2000Cattapan_61_067301,PLB2014SSZhang_730_30,PRC2015Xu_92_024324}, CSM~\cite{PRC2012QLiu_86_054312,PRC2014Shi_90_034319}, CMR~\cite{PRC2017Fang_95_024311}, CGF~\cite{PRC2016Shi_94_024302,EPJA2017Shi_53_40}, and RSM~\cite{NPA2004Hagino_735_55} methods have been developed to investigate resonances in deformed systems.

The Green's function (GF) method~\cite{PRB1992Tamura_45_3271,PRA2004Foulis_70_022706,Book2006Eleftherios-GF} is also an efficient tool for studying the single-particle resonant states. It has the following advantages: it treats the discrete bound states and the continuum on the same footing, it provides both the energies and the widths for the resonant states directly, and it produces the correct asymptotic behavior for the wave functions. Nonrelativistically and relativistically, there are already many applications of the GF method in nuclear physics to study the contribution of the continuum to the ground states and excited states. Nonrelativistically, in 1987, Belyaev \emph{et~al.} constructed, in the spherical case, the Green's function in Hartree-Fock-Bogoliubov (HFB) theory in the coordinate representation~\cite{SJNP1987Belyaev_45_783}. Afterwards, Matsuo applied this Green's function to the quasiparticle random-phase approximation (QRPA)~\cite{NPA2001Matsuo_696_371}, which was further used to describe the collective excitations coupled to the continuum~\cite{PTPS2002Matsuo_146_110, PRC2005Matsuo_71_064326,NPA2007Matsuo_788_307,PTP2009Serizawa_121_97,PRC2009Mizuyama_79_024313,PRC2010Matsuo_82_024318,PRC2011Shimoyama_84_044317}, microscopic structures of monopole pair vibrational modes and associated two-neutron transfer amplitudes in neutron-rich Sn isotopes~\cite{PRC2013Shimoyama_88_054308}, and neutron capture reactions in neutron-rich nuclei~\cite{PRC2015Matsuo_91_034604}. Recently, Zhang \emph{et~al.} developed the fully self-consistent continuum Skyrme-HFB theory using the GF method~\cite{PRC2011ZhangY_83_054301,PRC2012YZhang_86_054318,PRC2019XYQu_99_014314}, which was further extended for odd-A nuclei~\cite{PRC2019Sun}. In the deformed case, in 2009, Oba and Matsuo extended the continuum HFB theory to include deformation on the basis of a coupled-channel representation and explored the properties of the continuum and pairing correlation in deformed nuclei near the neutron drip line~\cite{PRC2009Oba_80_024301}. Relativistically, in the spherical case, in Refs.~\cite{PRC2009Daoutidis_80_024309,PRC2010DYang_82_054305}, the fully self-consistent relativistic continuum random-phase-approximation (RCRPA) was developed with the Green's function of the Dirac equation and used to study the contribution of the continuum to nuclear collective excitations. During the past decades, covariant density functional theory (CDFT) was very successful in describing nuclear structures~\cite{ANP1986Serot_16_1, PPNP1996PRing_37_193, JPG2015JMeng_42_093101,PRC2017Zhang_96_054308,PRC2018Zhang_97_054302,PRC2016TTSun,JPG2017Lu_44_125104,PRC2017TTSun_96_044312,CPC2018Sun,
PRC2018Liu,PRCSun_99_034310, Niu2009PRC}. Those successes urged the authors to develop continuum CDFT based on the GF method and to calculate with this method for the first time the energies and widths of single-particle resonant
states~\cite{PRC2014TTSun_90_054321,JPG2016TTSun_43_045107,PRC2017Ren_95_054318}.  In 2016 we included pairing correlations and developed Green's function relativistic continuum Hartree-Bogoliubov (GF-RCHB) theory, where the continuum was treated exactly~\cite{Sci2016Sun_46_12006}. It was applied for the study of giant halo phenomena in neutron-rich Zr isotopes.

In this work, we aim to develop a self-consistent deformed continuum covariant density functional theory using the GF method and treating pairing correlations, deformations, and the continuum on the same footing. As a first step, the GF method for a deformed Dirac equation, {\it i.e.}, a deformed GF-Dirac model, will be implemented and applied to an illustrative calculation of single-neutron bound and resonant states in relativistic quadrupole-deformed Woods-Saxon potentials. The paper is organized as follows. In Sec.~\ref{sec:form}, we give the formulation of the coupled-channel Dirac equation and the Green's function partial-wave expansion with exact boundary conditions. After the numerical details in Sec.~\ref{sec:Numer}, we present results and discussions in Sec.~\ref{sec:resu}. Finally, a brief summary and some perspectives are given in Sec.~\ref{sec:Sum}.


\section{\label{sec:form}Formalism}

\subsection{Coupled-channel Dirac equation}

In the CDFT~\cite{ANP1986Serot_16_1, RPP1989Reinhard_52_439,PPNP1996PRing_37_193, PR2005Vretenar_409_101, PPNP2006MengJ_57_470,JPG2015JMeng_42_093101}, nucleons are described as Dirac spinors moving in a mean-field potential, and the corresponding Dirac equation is
\begin{equation}
[{\bm \alpha}\cdot{\bm p}+V({\bm r})+\gamma_0(M+S({\bm r}))]\psi({\bm r})=\varepsilon\psi({\bm r}),
\label{Eq:DiracEq}
\end{equation}
where ${\bm \alpha}$ and $\gamma_0$ are Dirac matrices, $M$ is the nucleon mass, and $S({\bm r})$ and $V({\bm r})$ are the scalar and vector potentials, respectively. In the following discussions, for simplicity, only the axially quadrupole deformations are considered and the potentials take the following form,
\begin{subequations}
\begin{eqnarray}
&&S({\bm r})=S_0(r)+S_2(r)Y_{20}(\theta,\phi),\\
&&V({\bm r})=V_0(r)+V_2(r)Y_{20}(\theta,\phi),
\end{eqnarray}%
\label{Eq:Potential}%
\end{subequations}%
where $S_0(r)$ and $V_0(r)$ are the spherical parts and $S_2(r)Y_{20}(\theta,\phi)$ and $V_2(r)Y_{20}(\theta,\phi)$ are the quadrupole parts.

For a nucleon in an axially quadrupole-deformed potential, parity $\pi$ and the $z$-component $\Omega$ of the angular momentum are good quantum numbers and the single-particle wave function can be expanded in terms of spherical Dirac spinors,
\begin{equation}
  \psi_\Omega=\sum_{\kappa}
  \left(
  \begin{array}{c}
  {\displaystyle i \frac{G_{\Omega \kappa}(r)}{r}}\\
  {\displaystyle \frac{F_{\Omega \kappa}(r)}{r}{\sigma\cdot\hat {\bm r}}}
  \end{array}
  \right)
  Y_{\kappa\Omega}(\theta,\phi),%
\label{Eq:Dspinor}%
\end{equation}%
where $G_{\Omega \kappa}(r)/r$ and $F_{\Omega \kappa}(r)/r$ are, respectively, the radial wave functions for the upper and lower components, $Y_{\kappa\Omega}(\theta,\phi)$ are the spinor spherical harmonics~\cite{Book1988Varshalovich-QuantumTheory}, and the quantum number $\kappa$ is related with the angular momentums $l$ and $j$,
\begin{equation}
\left\{
  \begin{array}{ll}
    l=\kappa,j=\kappa-\frac{1}{2}, & \hbox{if $\kappa>0$;} \\
    l=-\kappa-1,j=-\kappa-\frac{1}{2}, & \hbox{if $\kappa<0$,}
  \end{array}
\right.
\end{equation}
which labels different ``channels".
The Dirac equation~(\ref{Eq:DiracEq}) is then transformed to a coupled-channel form for the radial wave functions,
\begin{subequations}
\begin{eqnarray}
    \label{Eq:G}
    &&\frac{{\rm d}G_{\Omega \kappa}}{{\rm d}r}+\frac{\kappa}{r}G_{\Omega \kappa}-(\varepsilon_\Omega+2M)F_{\Omega \kappa}\nonumber\\
    &&~~~~~~+\sum_{\kappa'\lambda}(V_\lambda-S_\lambda)A(\lambda,\kappa',\kappa,\Omega)F_{\Omega \kappa'}=0,\\
    \label{Eq:F}
    &&\frac{{\rm d}F_{\Omega \kappa}}{{\rm d}r}-\frac{\kappa}{r}F_{\Omega \kappa}+\varepsilon_\Omega G_{\Omega \kappa}
    \nonumber\\
    &&~~~~~~-\sum_{\kappa'\lambda}(V_\lambda+S_\lambda)A(\lambda,\kappa',\kappa,\Omega)G_{\Omega \kappa'}=0,%
\end{eqnarray}%
\label{Eq:DDirac}%
\end{subequations}%
where the index $\lambda$ of the potentials is $0$ and $2$, respectively, for the spherical and quadrupole parts of the potentials, and $A(\lambda,\kappa',\kappa,\Omega)$ has the form,
\begin{eqnarray}
&&A(\lambda,\kappa',\kappa,\Omega)=
\langle Y_{\kappa\Omega}|Y_{\lambda 0}|Y_{\kappa'\Omega}\rangle\nonumber\\
&&\hspace{0.1cm}=(-1)^{\Omega+\frac{1}{2}}\frac{\hat{j}\hat{j}'}{\sqrt{4\pi}}
                               \left(
                                \begin{array}{ccc}
                                    j & \lambda & j'\\
                                    -\Omega       & 0 & \Omega
                                \end{array}
                                \right)
                                \left(
                                \begin{array}{ccc}
                                   j'  & \lambda & j\\
                                    \frac{1}{2}  & 0 & -\frac{1}{2}
                                \end{array}
                                \right).
                                \label{Eq:A-CG}
\end{eqnarray}%
with $\hat{j}=\sqrt{2j+1}$. Note that the single-particle energy in Eq.~(\ref{Eq:DDirac}) is shifted by $M$ with respect to that in Eq.~(\ref{Eq:DiracEq}). The couplings among different channels in Eq.~(\ref{Eq:DDirac}) are governed by the deformed potentials,
\begin{equation}
v^{\pm}_{\kappa\kappa'}=\sum_{\lambda}(V_\lambda\pm S_\lambda)A(\lambda,\kappa',\kappa,\Omega).
\end{equation}
In the practical calculations, we have to truncate the partial-wave expansion and we denote $N$ to represent the number of partial waves to be included. Note that in Refs.~\cite{PRC1987Price,PRC2010LiZP_81_034311,PRC2015Xu_92_024324}, a coupled-channel Dirac equation is also solved in the coordinate space for deformed nuclei.

\subsection{Boundary conditions for the wave functions}

To describe the single-particle states properly, correct asymptotic boundary conditions have to be considered at $r\rightarrow 0$ and $r\rightarrow\infty$.

At $r\rightarrow 0$, the Dirac spinor is regular and satisfies
\begin{eqnarray}
 \left(
   \begin{array}{c}
     G_{\Omega \kappa}(r) \\
     F_{\Omega \kappa}(r) \\
   \end{array}
 \right)
   &\longrightarrow& r\left(
                                                       \begin{array}{c}
                                                         j_l(k r) \\
                                                         \frac{\kappa}{|\kappa|}\frac{\varepsilon-v^{+}_{\kappa\kappa'}}{k}j_{\tilde{l}}(kr)\\
                                                       \end{array}
                                                     \right),
                                                  \label{Eq:behavior_r0}
 \end{eqnarray}
where the quantum number $\tilde{l}$ is defined as $\tilde{l}=l+(-1)^{j+l-1/2}$, $k=\sqrt{(\varepsilon-v^{+}_{\kappa\kappa'})(\varepsilon-v^{-}_{\kappa\kappa'}+2M)}$ for all states, and $j_l(k r)$ is the spherical Bessel function of the first kind satisfying
\begin{equation}
j_l(k r)\longrightarrow \frac{(kr)^l}{(2l+1)!!},~~~\text{ when}~~r\rightarrow 0.
\end{equation}
Note that the single-particle energy $\varepsilon$ in the whole paper is a complex number if without a special request.

At $r\rightarrow\infty$, the Dirac spinor is exponentially decaying for the bound states and oscillating outgoing for the continuum. We have
\begin{eqnarray}
 \left(
   \begin{array}{c}
     G_{\Omega \kappa}(r) \\
     F_{\Omega \kappa}(r) \\
   \end{array}
 \right)
 &\longrightarrow&\left(
                                                    \begin{array}{c}
                                                      rk h^{(1)}_l(k r) \\
                                                      \frac{\kappa}{|\kappa|}\frac{rk^2}{\varepsilon+2M}h^{(1)}_{\tilde{l}}(k r) \\
                                                    \end{array}
                                                  \right),
\label{Eq:behavior_rinf}
\end{eqnarray}
where $h^{(1)}_l(k r)$ is the spherical Hankel function of the first kind, and $k=\sqrt{\varepsilon(\varepsilon+2M)}$.

\subsection{Green's function partial-wave expansion}

The Green's function defined for the Dirac equation in the coordinate space obeys
\begin{equation}
[\varepsilon-\hat{h}({\bm r})]\mathcal{G}({\bm r},{\bm r'};\varepsilon)=\delta({\bm r}-{\bm r'}),
\label{Eq:GF-definition}
\end{equation}
where $\hat{h}({\bm r})$ is the Dirac Hamiltonian and $\varepsilon$ is an arbitrary single-particle energy. With a complete set of eigenstates $\psi_n({\bm r})$~($n=\Omega^\pi$) and eigenvalues $\varepsilon_n$ of the Dirac equation, the Green's function can also be represented as
\begin{equation}
\mathcal{G}({\bm r},{\bm r'};\varepsilon)=\sum_n\frac{\psi_n({\bm r})\psi_{n}^\dag({\bm r'})}{\varepsilon-\varepsilon_n}.
\label{Eq:GF}
\end{equation}
Corresponding to the upper and lower components of the Dirac spinors $\psi_{n}({\bm r})$, the Green's function assumes a $2\times 2$ matrix form,
\begin{equation}
\mathcal{G}({\bm r},{\bm r'};\varepsilon)=
\left(
  \begin{array}{cc}
    \mathcal{G}^{(11)}({\bm r},{\bm r'};\varepsilon) & \mathcal{G}^{(12)}({\bm r},{\bm r'};\varepsilon) \\
    \mathcal{G}^{(21)}({\bm r},{\bm r'};\varepsilon) & \mathcal{G}^{(22)}({\bm r},{\bm r'};\varepsilon) \\
  \end{array}
\right).
\end{equation}

Using the partial-wave expansion, the Green's function with a given $\Omega$ can be expanded as
\begin{equation}
\mathcal{G}_{\Omega}({\bm r},{\bm r'};\varepsilon)=\sum_{\kappa\kappa'}Y_{\kappa\Omega}(\theta,\phi)\frac{\mathcal{G}_{\Omega \kappa\kappa'}(r,r';\varepsilon)}{rr'}Y_{\kappa'\Omega }^{*}(\theta',\phi'),
\label{Eq:GF-expansion}
\end{equation}
where $\mathcal{G}_{\Omega \kappa\kappa'}(r,r';\varepsilon)$ is the radial Green's function coupling the partial waves $\kappa$ and $\kappa'$ and of $2N\times2N$ matrix form,
\begin{equation}
\mathcal{G}_{\Omega \kappa\kappa'}(r,r';\varepsilon)
=
\left(
  \begin{array}{cc}
    \mathcal{G}_{~\Omega \kappa\kappa'}^{(11)}(r,r';\varepsilon) & \mathcal{G}_{~\Omega \kappa\kappa'}^{(12)}(r,r';\varepsilon) \\
    \mathcal{G}_{~\Omega \kappa\kappa'}^{(21)}(r,r';\varepsilon) & \mathcal{G}_{~\Omega \kappa\kappa'}^{(22)}(r,r';\varepsilon) \\
  \end{array}
\right),
\end{equation}
where $N$ is the number of partial waves under consideration.

According to the definition of the Green's function for the Dirac equation in Eq.~(\ref{Eq:GF-definition}), it can be easily derived that the radial Green's function $\mathcal{G}_{\Omega \kappa\kappa'}(r,r';\varepsilon)$ satisfies the following coupled-channel equation,
\begin{eqnarray}
&&\hspace{-0.5cm}\left(
  \begin{array}{cc}
    -\varepsilon & {\displaystyle -\frac{d}{dr}+\frac{\kappa}{r}} \\
   {\displaystyle \frac{d}{dr}+\frac{\kappa}{r}} & -\varepsilon-2M \\
  \end{array}
\right)\mathcal{G}_{\Omega \kappa\kappa'}(r,r';\varepsilon)\nonumber\\
&&\hspace{-0.5cm}+\sum_{ \kappa''}\left(
                                     \begin{array}{cc}
                                       {\displaystyle v^+_{\kappa\kappa''}} & 0 \\
                                       0 & {\displaystyle v^-_{\kappa\kappa''}} \\
                                     \end{array}
                                   \right)
\mathcal{G}_{\Omega \kappa''\kappa'}(r,r';\varepsilon)
=\frac{\delta(r-r')}{rr'}J,~~
\label{Eq:GF-partial}
\end{eqnarray}
where
\begin{equation}
J=\left(
  \begin{array}{cc}
    1 & 0 \\
    0 & -1 \\
  \end{array}
\right)\otimes I_{N},
\end{equation}
with $I_{N}$ being the $N$-dimensional unit matrix.

Finally, a Green's function considering the exact asymptotic behavior of the Dirac spinors and satisfying Eq.~(\ref{Eq:GF-partial}) can be constructed as
\begin{eqnarray}
&&\mathcal{G}_{\Omega \kappa\kappa'}(r,r';\varepsilon)=\nonumber\\
&&\sum_{\kappa''\kappa'''}\left[\phi^{({\rm in})*}_{\Omega \kappa''\kappa}(r,\varepsilon)W^{-1}_{ \Omega \kappa''\kappa'''}\phi^{({\rm out})}_{\Omega \kappa'''\kappa'}(r',
\varepsilon)\theta(r'-r)\right.\nonumber\\
&&\hspace{0.22cm}\left.+\phi^{({\rm out})*}_{\Omega \kappa''\kappa}(r,\varepsilon)W^{-1}_{\Omega \kappa'''\kappa''}\phi^{({\rm in})}_{\Omega \kappa'''\kappa'}(r',\varepsilon)\theta(r-r')\right],%
\label{Eq:GF_element}%
\end{eqnarray}%
where $\theta(r-r')$ is the radial step function, $\phi^{({\rm in})}_{\Omega \kappa\kappa''}(r, \varepsilon)$ and $\phi^{({\rm out})}_{\Omega \kappa\kappa''}(r, \varepsilon)$ are two linearly independent Dirac spinors
\begin{eqnarray}
\phi^{({\rm in})}_{\Omega \kappa\kappa''}(r,\varepsilon)&=&\left(
                            \begin{array}{c}
                              G^{({\rm in})}_{\Omega \kappa\kappa''}(r,\varepsilon) \\
                              F^{({\rm in})}_{\Omega \kappa\kappa''}(r,\varepsilon) \\
                            \end{array}
                          \right),\nonumber\\
\phi^{({\rm out})}_{\Omega \kappa\kappa''}(r,\varepsilon)&=&\left(
                            \begin{array}{c}
                              G^{({\rm out})}_{\Omega \kappa\kappa''}(r,\varepsilon) \\
                              F^{({\rm out})}_{\Omega \kappa\kappa''}(r,\varepsilon) \\
                            \end{array}
                          \right),
\end{eqnarray}
which are obtained by integrating the following coupled-channel Dirac equation
\begin{eqnarray}
&&\hspace{-0.5cm}\left(
  \begin{array}{cc}
    -\varepsilon & {\displaystyle -\frac{d}{dr}+\frac{\kappa}{r}} \\
   {\displaystyle \frac{d}{dr}+\frac{\kappa}{r}} & -\varepsilon-2M \\
  \end{array}
\right)\left(
         \begin{array}{c}
           G_{\Omega \kappa\kappa''}(r, \varepsilon) \\
           F_{\Omega \kappa\kappa''}(r, \varepsilon) \\
         \end{array}
       \right) \nonumber \\
&&\hspace{-0.7cm}     +\sum_{\kappa'}\left(
                   \begin{array}{cc}
                     {\displaystyle v^+_{\kappa\kappa'}} & 0 \\
                     0 & {\displaystyle v^-_{\kappa\kappa'}} \\
                   \end{array}
                 \right)\left(
                          \begin{array}{c}
                            G_{\Omega \kappa'\kappa''}(r, \varepsilon) \\
                            F_{\Omega \kappa'\kappa''}(r, \varepsilon) \\
                          \end{array}
                        \right)=0,%
\end{eqnarray}%
in $r$ space using a fourth-order Runge-Kutta algorithm starting from the boundary conditions at $r\rightarrow 0$ and $r\rightarrow \infty$, respectively. The Dirac spinor matrices $G_\Omega^{\rm (in/out)}$ and $F_\Omega^{\rm (in/out)}$ take the following form at the boundaries
\begin{equation}
\hspace{-0.5cm}
\left(
  \begin{array}{c}
    G^{\rm (in/out)}_{\Omega} \\
     F^{\rm (in/out)}_{\Omega} \\
  \end{array}
\right)
\rightarrow\left(
                         \begin{array}{cccc}
                            G^{\rm (in/out)}_{\Omega \kappa_1\kappa_1} & 0 & \cdots & 0 \\
                           0 & G^{\rm (in/out)}_{\Omega \kappa_2\kappa_2} & \cdots & 0 \\
                           \vdots & \vdots & \ddots & \vdots \\
                           0 & 0 & \cdots &G^{\rm (in/out)}_{\Omega \kappa_N\kappa_N} \\
                              F^{\rm (in/out)}_{\Omega \kappa_1\kappa_1} & 0 & \cdots & 0 \\
                           0 & F^{\rm (in/out)}_{\Omega \kappa_2\kappa_2} & \cdots & 0 \\
                           \vdots & \vdots & \ddots & \vdots \\
                           0 & 0 & \cdots &F^{\rm (in/out)}_{\Omega \kappa_N\kappa_N} \\
                         \end{array}
                       \right),
\end{equation}
where $G^{\rm (in)}_{\Omega \kappa\kappa}$ and $F^{\rm (in)}_{\Omega \kappa\kappa}$ are the asymptotic solutions $G_{\Omega\kappa}(r)$ and $F_{\Omega\kappa}(r)$ at $r\to 0$ in Eq.~(\ref{Eq:behavior_r0}) and $G^{\rm (out)}_{\Omega \kappa\kappa}$ and $F^{\rm (out)}_{\Omega \kappa\kappa}$ are those at $r\to\infty$ in Eq.~(\ref{Eq:behavior_rinf}). The subscript $\kappa_{i}$ denotes the value of $\kappa$ in the $i_\text{th}$ channel.

In Eq.~(\ref{Eq:GF_element}), $W_{\Omega \kappa''\kappa'''}$ is the Wronskian matrix element
\begin{eqnarray}
&&W_{\Omega \kappa''\kappa'''}=\sum_{\kappa}\left[G^{({\rm out})}_{\Omega \kappa''\kappa}(r,\varepsilon)F^{({\rm in})*}_{\Omega \kappa'''\kappa}(r,\varepsilon)\right.\nonumber\\
&&\hspace{2.3cm}\left.-F^{({\rm out})}_{\Omega \kappa''\kappa}(r,\varepsilon)G^{({\rm in})*}_{\Omega \kappa'''\kappa}(r,\varepsilon)\right].
\end{eqnarray}
It does not depend on $r$.

\subsection{Density of states}

Following Ref.~\cite{Book2006Eleftherios-GF}, we can study the single-particle spectrum by calculating the density of states (DOS) $n(\varepsilon)$,
\begin{equation}
n(\varepsilon)=\sum_n\delta(\varepsilon-\varepsilon_n),
\label{Eq:dos1}
\end{equation}
where $\varepsilon_{n}$ is the eigenvalue of the Dirac equation, $\varepsilon$ is a real single-particle energy for the bound states and physical continuum, and $\sum_n$ means a summation over the discrete states and an integration for the physical continuum. For the bound states, the DOS $n(\varepsilon)$ exhibits discrete $\delta$-functions at $\varepsilon=\varepsilon_{n}$, while in the physical continuum it has a continuous distribution.

By introducing an infinitesimal imaginary part $``i\epsilon"$ in the energy $\varepsilon$, it can be proven that $n(\varepsilon)$ can be calculated by integrating the imaginary part of the Green's function over the coordinate space~\cite{PRC2014TTSun_90_054321},
\begin{equation}
n(\varepsilon)=-\frac{1}{\pi }\int d{\bm r}{\rm Im}[\mathcal{G}^{(11)}({\bm r},{\bm r};\varepsilon+i\epsilon)
+\mathcal{G}^{(22)}({\bm r},{\bm r};\varepsilon+i\epsilon)].
\label{Eq:DOSGF}
\end{equation}
The density of states for each value of $\Omega^\pi$ is
\begin{eqnarray}
&&n_{\Omega}(\varepsilon)=-\frac{2}{\pi }\sum_\kappa\int d{r}{\rm Im}[\mathcal{G}_{\Omega \kappa\kappa}^{(11)}({r},{r};\varepsilon+i\epsilon)\nonumber \\
&&\hspace{3.3cm}+\mathcal{G}_{\Omega \kappa\kappa}^{(22)}({r},{ r};\varepsilon+i\epsilon)].
\label{EQ:DOS}
\end{eqnarray}
Note that with the infinitesimal imaginary part $``i\epsilon"$ in the single-particle energy, the density of states for discrete single-particle states in form of $\delta$-functions~(no width) are simulated by a Lorentzian functions with the full-width at half-maximum (FWHM) of $2\epsilon$.

\section{Numerical details}\label{sec:Numer}

In this work, the radial parts of the quadrupole-deformed potentials in the Dirac equation~(\ref{Eq:DiracEq}) are taken in a Woods-Saxon form as follows,
\begin{eqnarray}
S_{0}(r)&=&S_{\rm WS}f(r),\nonumber\\
V_{0}(r)&=&V_{\rm WS}f(r),\nonumber\\
S_{2}(r)&=&-\beta S_{\rm WS}k(r),\nonumber\\
V_{2}(r)&=&-\beta V_{\rm WS}k(r),
\end{eqnarray}
with
\begin{equation}
{\displaystyle f(r)=\frac{1}{1+{\rm{exp}}(\frac{r-R}{a})}~~\text{and}~~k(r)=r\frac{df(r)}{dr}}.
\end{equation}

To compare the present results with those obtained by the CSM, ACCC, and SPS methods, we adopt the same parameters for the potentials as in Ref.~\cite{PRC2015Xu_92_024324}, which are determined by reproducing the results for the $p$-wave halo candidate nucleus $^{37}$Mg~\cite{PRL2014Kobayashi_112_242501} by self-consistent spherical RHB theory with the PC-PK1 parameter set~\cite{PRC2010Zhao_82_054319}.
Specifically, the depths of the scalar and vector potentials are chosen as $S_{\rm WS}=-420.3$~MeV and $V_{\rm WS}=349.7$~MeV, respectively, the radius $R=3.705$~fm, the diffuseness $a=0.67$~fm, and $\beta$ is the axial deformation parameter of the potential.

The coupled-channel Dirac equation is solved in the radial space with a mesh step of $0.1$~fm and a cutoff at $R_{\rm box}=20$~fm.  To calculate the density of states $n_{\Omega}(\varepsilon)$, the parameter $\epsilon$ in Eq.~(\ref{EQ:DOS}) is taken as $1\times10^{-6}$~MeV and the energy step $d\varepsilon$ is $1\times10^{-4}$~MeV. With this energy step, the accuracy for energies and widths of the single-particle resonant states can be up to $0.1$~keV. Furthermore, a higher degree of accuracy can be achieved for energies and widths if we take smaller energy steps $d\varepsilon$.

\section{\label{sec:resu}Results and discussion}

\begin{figure}[!t]
\includegraphics[width=\linewidth]{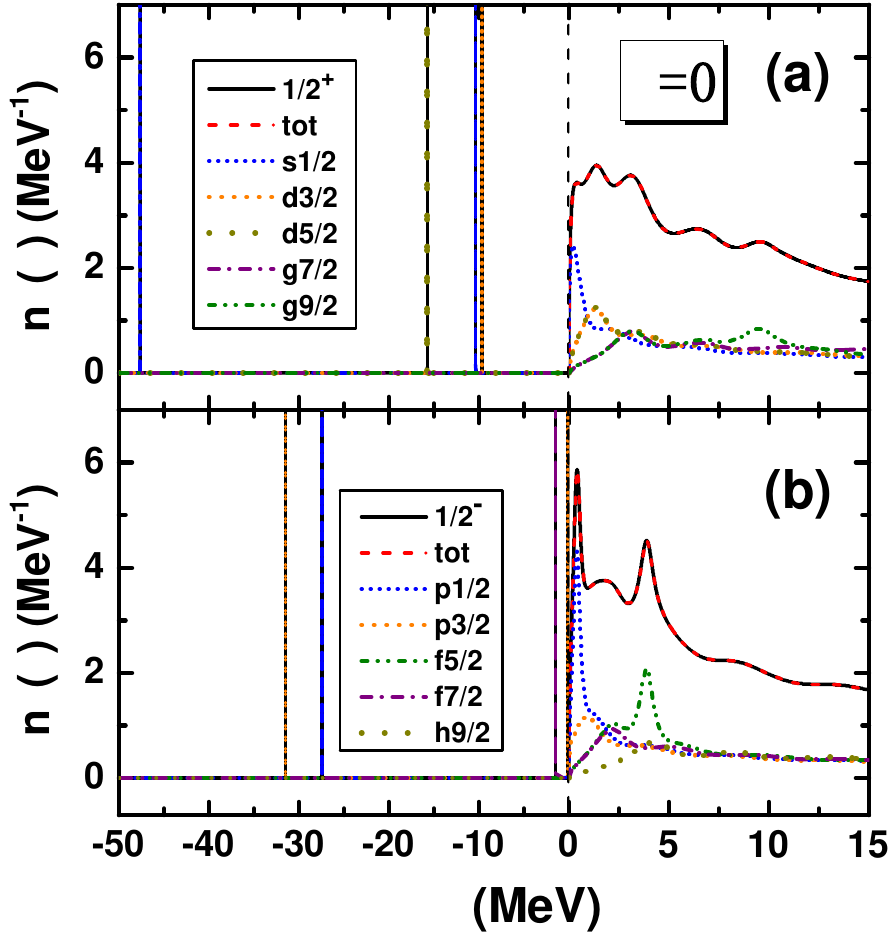}
\caption{\label{fig1}(Color online) Density of states $n_{\Omega}(\varepsilon)$ for neutrons with $\Omega^{\pi}=1/2^+$ (a) and $\Omega^{\pi}=1/2^-$ (b), obtained by solving the coupled-channel Dirac equations using the GF method by setting the deformation parameter $\beta=0$ and the number of coupled partial waves $N=5$~(black-solid lines), in comparison with the results calculated from the spherical GF code (colored-dashed lines). The red-dashed line in the continuum is the total density of states by summing those for five spherical states.}
\end{figure}

\begin{figure}[!t]
\includegraphics[width=\linewidth]{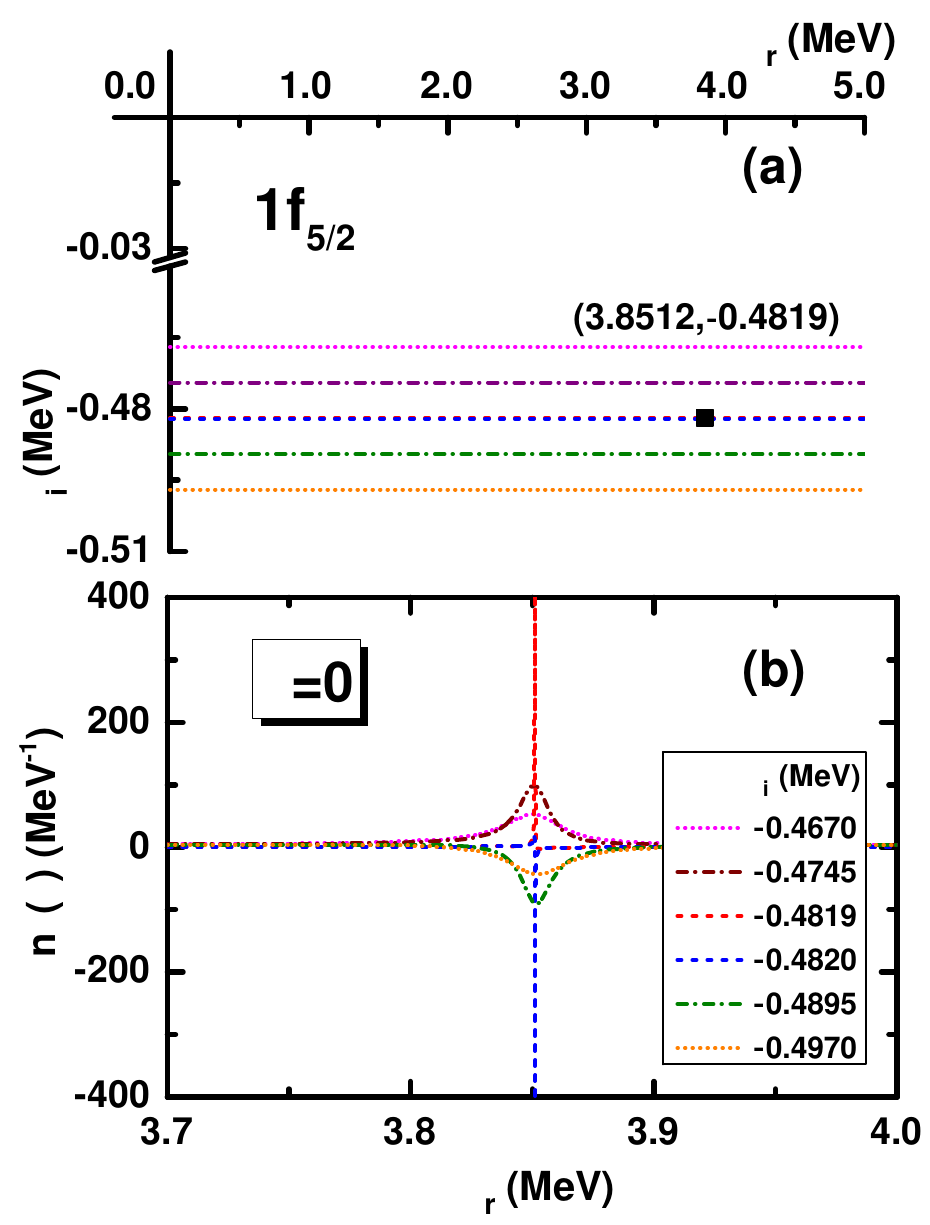}
\caption{\label{fig2}(Color online) (a)~The single-particle resonant state $1f_{5/2}$ in the fourth quadrant of the complex energy panel.
(b)~Densities of states $n_{\Omega}(\varepsilon)$ as functions of the real part of the complex energy $\varepsilon_{r}$ for different values of the imaginary part $\varepsilon_{i}$, obtained by solving the coupled-channel Dirac equations using the GF method by setting $\Omega^{\pi}=5/2^-$, the number of coupled partial waves $N=1$, and the deformation parameter $\beta=0$.}
\end{figure}

Firstly, a numerical check for the new developed deformed GF-Dirac code is carried out by comparing the results calculated from the coupled-channel Dirac equation by setting the deformation parameter $\beta=0$ with those from the spherical GF code~\cite{PRC2014TTSun_90_054321}. In Fig.~\ref{fig1}, the density of states $n_{\Omega}(\varepsilon)$ for neutrons with $\Omega^{\pi}=1/2^+$ and $1/2^-$ are plotted as a function of the single-particle energy $\varepsilon$. The peaks of  $\delta$-functional shape below the continuum threshold correspond to the single-particle bound states and the spectra with $\varepsilon>0$ describe the continuum. In each panel, the black-solid line denotes the results obtained by solving the coupled-channel Dirac equation with $N=5$ coupled partial waves,  while the colored-dashed lines are those by solving the spherical radial Dirac equation. In the continuum, the red-dashed line is the total density of states calculated by summing those of corresponding five states in the spherical Dirac equation. In panel~(a), the five partial waves for the state $\Omega^{\pi}=1/2^+$ are chosen as $s_{1/2}$, $d_{3/2}$, $d_{5/2}$, $g_{7/2}$, and $g_{9/2}$, and in panel~(b), those for $\Omega^{\pi}=1/2^-$ are $p_{1/2}$, $p_{3/2}$, $f_{5/2}$, $f_{7/2}$, and $h_{9/2}$. Note that a factor of $2/(2j+1)$ is multiplied in calculating the density of states for the corresponding spherical state in the spherical code, considering the difference of the degree of degeneracy between the deformed code and spherical one. It is clearly shown that all the peaks below the threshold and spectrum in the continuum obtained by the deformed and spherical codes are completely consistent.

From the density of states, the energies for the bound states and the energies and widths for the resonant states can be extracted. In our previous works~\cite{PRC2014TTSun_90_054321,JPG2016TTSun_43_045107,PRC2017Ren_95_054318}, the resonant states were identified by comparing the density of states for nucleons in the potentials $V(r)$  and $S(r)$ with those for free particles obtained with the potentials $V(r)=S(r)=0$~(see Fig.~6 in Ref.~\cite{PRC2014TTSun_90_054321} as an example). The resonance energy and width, in this framework, are defined as the positions and the FWHM of the resonant peaks, which are the DOS differences between nucleons in the mean-field potential and free particles. This approach based on the GF method can describe narrow resonant states very well, while its accuracy decreases for wide resonances.

It is well known that the resonant states are the poles in the fourth quadrant of complex energy panel. Therefore, here we propose a direct approach that is to search for the poles according to the extremes of the GF (c.f.~Eq.(\ref{Eq:GF})) by scanning the complex energy $\varepsilon$, both the real part $\varepsilon_{r}$ and the imaginary part $\varepsilon_i$, in the fourth quadrant. In practice, we will still apply the definition DOS in Eqs.~(\ref{Eq:DOSGF}) and (\ref{EQ:DOS}) to find the poles or extremes. The infinitesimal imaginary part $``i\epsilon"$ is taken to be zero as the single-particle energy $\varepsilon$ in the equations is a complex number. As an example, in Fig.~\ref{fig2} we show the way to determine the spherical resonant state $1f_{5/2}$. The coupled-channel Dirac equation is solved using the GF method by setting $\Omega^{\pi}=5/2^-$, the number of coupled partial waves is $N=1$, and the deformation parameter is $\beta=0$. As shown in Fig.~\ref{fig2}(a), a wide range of complex energies $\varepsilon$ including the real part $\varepsilon_{r}$ and the imaginary part $\varepsilon_{i}$ are scanned to calculate the DOSs to search for the pole. In Fig.~\ref{fig2}(b), we plot the corresponding DOSs as functions of $\varepsilon_{r}$ for different $\varepsilon_i$. When the imaginary energy $\varepsilon_i$ varies from $-0.467~$MeV to $-0.497~$MeV, the DOSs change dramatically in the energy range between $\varepsilon_r=3.8~$MeV and $3.9~$MeV. As the absolute value $|\varepsilon_i|$ increases, the peak of DOS becomes higher and higher and reaches an extreme when $\varepsilon_i=-0.4819~$MeV and the peak locates at $\varepsilon_{r}=3.8512$~MeV. One notes a sharp reversal of the DOS at the next energy point $\varepsilon_i=0.4820~$MeV and the peak of the DOS becomes lower and lower when $|\varepsilon_i|$ increases further. This indicates that there is a pole at $\varepsilon=3.8512-i0.4819~$MeV. Hereafter, we also write the energy for resonant states in the form $\varepsilon=E-i\Gamma/2$, where $E$ and $\Gamma$ are the resonant energy and width, respectively.

\begin{table}[t!]
\center
\caption{Energies $E-i\Gamma/2$ (in MeV) for the single-neutron resonant states $2p_{1/2}$, $1f_{5/2}$, and $1g_{9/2}$, extracted from the density of states by the GF method as shown in Fig.~\ref{fig2}. For comparison, the results obtained by the CSM method are also listed. The deformation parameter $\beta=0$ and the number of coupled partial waves $N=1$.}
\label{Tab1}
\begin{tabular}{ccccc}
  \hline\hline
 $nl_{j}$   &&GF               &&CSM~\cite{Guo_CMR}\\ \hline
 $2p_{1/2}$ &&$0.3427-i0.2537$ &&$0.3427-i0.2537$\\
 $1f_{5/2}$ &&$3.8512-i0.4819$ &&$3.8512-i0.4819$\\
 $1g_{9/2}$ &&$9.4863-i1.6715$ &&$9.4864-i1.6715$\\
\hline\hline
\end{tabular}
\end{table}

In Table~\ref{Tab1}, we list the energies $E-i\Gamma/2$ for the resonant states $2p_{1/2}$, $1f_{5/2}$, and $1g_{9/2}$, obtained by solving the coupled-channel Dirac equation by the GF method, in comparison with the results obtained in the CSM approach, which has been proven to be very effective for both narrow and broad resonances~\cite{PRL2016Li_117_062502}. The deformation parameter is $\beta=0$ and the number of coupled partial waves is $N=1$. From Table~\ref{Tab1}, it can be seen that all the energies and widths obtained by the two methods are exactly equal, which is amazing since the the GF and CSM methods are two totally different methods. In the GF method, resonate states are obtained by solving the coupled-channel Dirac equation in coordinate space while in the CMR method, they are obtained by diagonalizing the Dirac Hamiltonian in momentum space. 
Compared to CMR method, GF method is more convenient to be used in combination with nuclear models with treating the bound and resonant states on the same footing.
In one word, the results in Fig.~\ref{fig1} and Table~\ref{Tab1} demonstrate that the present model is fully correct in the decoupled case with $\beta=0$.

\begin{figure}[tp!]
\includegraphics[width=\linewidth]{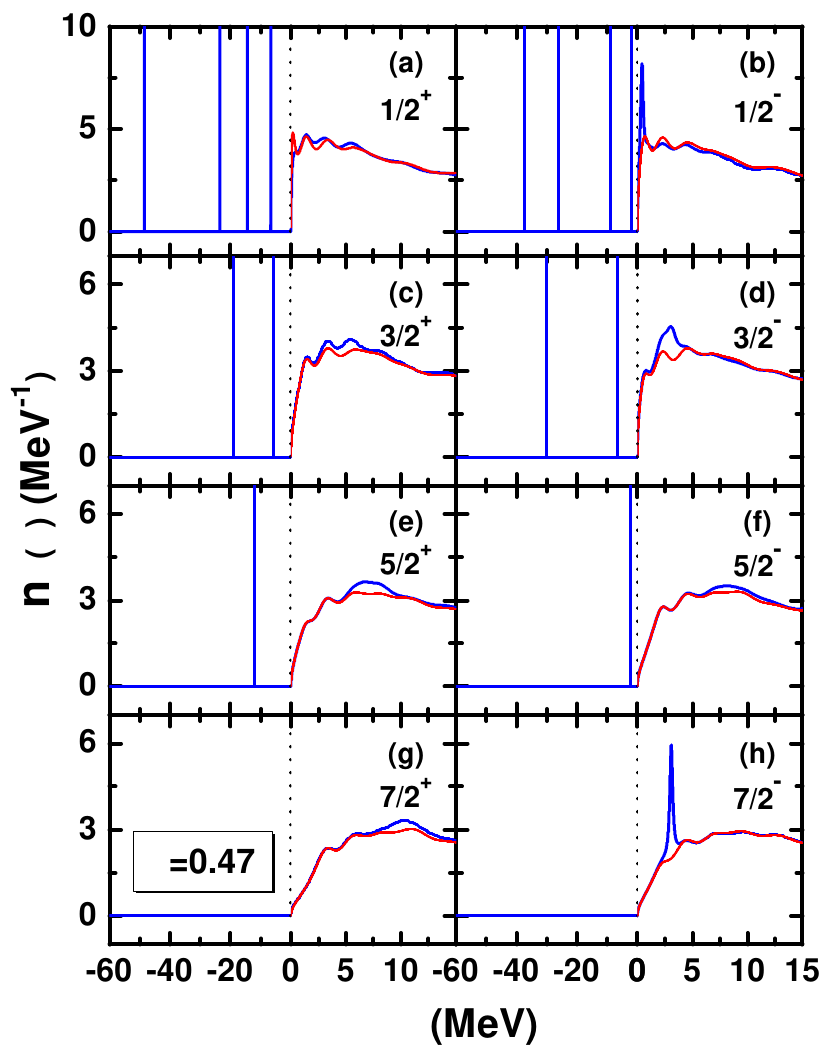}
\caption{\label{fig3}(Color online) Density of states $n_{\Omega}(\varepsilon)$ for neutrons with $\Omega^{\pi}=1/2^\pm, 3/2^\pm, 5/2^\pm$, and $7/2^\pm$ obtained by solving the coupled-channel Dirac equations with  quadrupole-deformed Woods-Saxon potentials using the GF method (blue-solid lines). The deformation parameter $\beta=0.47$ and the number of coupled partial waves $N=8$ are chosen in the calculations. Above the continuum threshold denoted by the black-dashed line, the results are compared with those for free neutrons calculated with the potentials $V(r)=S(r)=0$ (red-solid lines).}
\end{figure}

\begin{figure}[tp!]
\includegraphics[width=\linewidth]{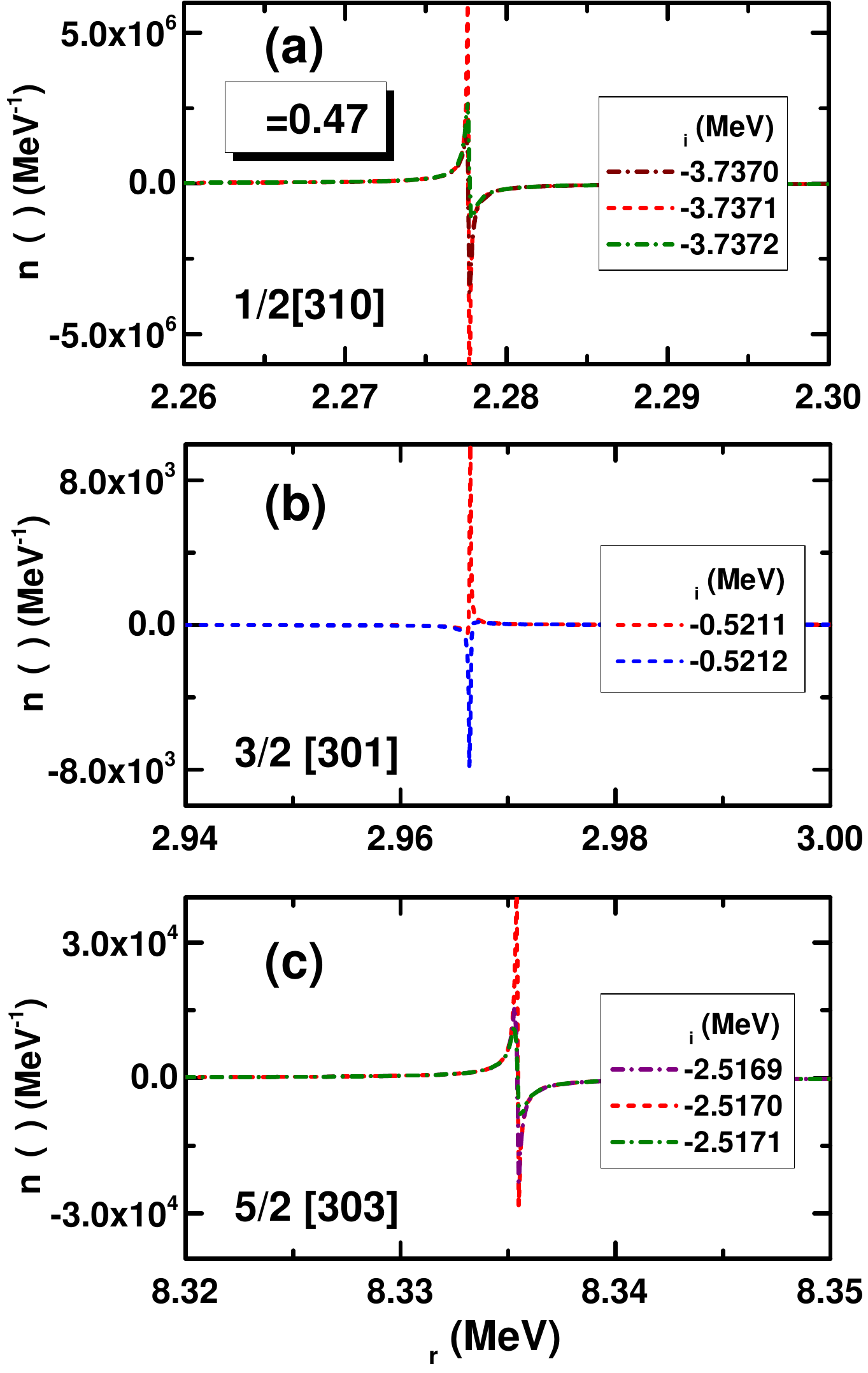}
\caption{\label{fig4}(Color online) Densities of states $n_{\Omega}(\varepsilon)$ for the resonant states $1/2[310]$, $3/2[301]$, and $5/2[303]$, as functions of the real part $\varepsilon_{r}$ of complex energy for different imaginary part $\varepsilon_{i}$. The Calculations are carried out by solving the coupled-channel Dirac equations with  quadrupole-deformed Woods-Saxon potentials with the GF method by choosing the deformation parameter $\beta=0.47$ and the number of coupled partial waves $N=8$.}
\end{figure}

Secondly, we further check the newly developed model by comparing the results with those from the CSM~\cite{PRC2017Fang_95_024311}, ACCC~\cite{PRC2015Xu_92_024324}, and SPS~\cite{PRC2010LiZP_81_034311} methods for a specific deformed potential with $\beta=0.47$ and the number of coupled partial waves $N=8$. In Fig.~\ref{fig3}, the density of states $n_{\Omega}(\varepsilon)$ for neutrons with $\Omega^{\pi}=1/2^\pm, 3/2^\pm, 5/2^\pm$, and $7/2^\pm$ are plotted as a function of single-neutron energy $\varepsilon$, obtained by solving the coupled-channel Dirac equation with a quadrupole-deformed potential. As in Fig.~\ref{fig1}, the peaks of $\delta$-functional shape below the continuum threshold correspond to bound states and the spectra with $\varepsilon>0$ are the continuum. Again, by comparing the density of states for bound neutrons (blue-solid lines) and those for free neutrons obtained with the potentials $V(r)=S(r)=0$ (red-solid lines) in the continuum, one can observe the resonant states in the $1/2^-$ and $7/2^-$ blocks, for which the density of states $n_{\Omega}(\varepsilon)$ sits on top of those for the free particles. However, for the other states, it is hard to determine the energies and widths with high precision due to the irregular shapes and the large widths. In addition, in the deformed case, one or more peaks may be observed for each $\Omega^\pi$ block due to the couplings of many channels. To obtain accurate energies and widths for those deformed resonant states, we carry out a similar analysis as in the spherical case shown in Fig.~\ref{fig2} by searching for the poles of the GF.

In Fig.~\ref{fig4}, taking the states $1/2[310]$, $3/2[301]$, and $5/2[303]$ as examples, which are the splitting states of the spherical resonant state $1f_{5/2}$, we plot the densities of states $n_{\Omega}(\varepsilon)$ by scanning the complex energy $\varepsilon$ in a wide range. The coupled-channel Dirac equation is solved by the GF method taking $\Omega^{\pi}=1/2^-,3/2^-$, and $5/2^-$, the deformation parameter $\beta=0.47$, and the number of coupled partial waves $N=8$. Convergence is checked for the number $N$ of coupled partial waves and it is found that $N=8$ is enough. Hereafter, the states are labeled by the Nilsson quantum numbers $\Omega[\mathcal{N}n_z\Lambda]$, where $\mathcal{N}$ is the principal quantum number, $n_z$ is the number of nodes of the wave function in the $z$ direction, and $\Lambda$ is the projection of the orbital angular momentum $l$ onto the $z$ axis.

For the three states in Fig.~\ref{fig4}, extremes are observed at $\varepsilon_r+i\varepsilon_i=2.2776-i3.7371$~MeV, $2.9665-i0.5211$~MeV, and $8.3354-i2.5170~$MeV, respectively, indicating that those are resonant states. We also note that the DOSs change signs following either the real part of the complex energy $\varepsilon_r$ or the imaginary part $\varepsilon_i$ when the DOSs are at their extremes. This is consistent with the fact that the resonances are poles according to Eq.~(\ref{Eq:GF}). 

\begin{table*}[tp!]
\center
\caption{Energies $E-i\Gamma/2$ (in MeV) of the single-neutron resonant states $\Omega[\mathcal{N}n_z\Lambda]$ obtained by solving the coupled-channel Dirac equations with quadrupole-deformed Woods-Saxon potentials using the GF method, in caparison with those by the CMR, ACCC and SPS approaches. The deformation parameter $\beta=0.47$ is chosen in all the calculations and the number of coupled partial waves $N=8$ is chosen in the GF, ACCC, and SPS calculations.}
\label{Tab2}
\begin{tabular}{ccccccc}
      \hline\hline								
	  positive parity                 &$1/2[440]$		&~$ 3/2[431]$		&~$5/2[422]$		&~$7/2[413]$	        &&  \\\hline		
      GF 	                          &$5.2908-i2.2471$	&~$5.0140-i2.0009$	&~$7.1199-i1.2230$	&~$10.2939-i1.9762$	&& \\	
      CMR~\cite{Guo_CMR}              &$5.2908-i2.2470$	&~$5.0140-i2.0009$	&~$7.1199-i1.2230$	&~$10.2939-i1.9762$	&&\\	
      ACCC~\cite{PRC2015Xu_92_024324} &$1.5015-i1.4018$	&~$4.1096-i1.5068$	&~$7.0017-i1.2653$	&~$10.1481-i1.7751$	&&\\		
      SPS~\cite{PRC2010LiZP_81_034311}&$1.7000-i1.7036$  &~$4.4800-i1.7937$	&~$7.0600-i1.2787$	&~$10.2000-i2.1367$	&&\\			
      \hline\hline
  	  negative parity                 &$1/2[301]$		&~$1/2[310]$        &~$3/2[312]$       &~$3/2[301]$		    &~$5/2[303]$		&~$7/2[303]$	\\\hline
      GF 	                          &$0.3865-i0.1674$  &~$2.2776-i3.7371$  &~$1.3190-i2.7176$ &~$2.9665-i0.5211$   &~$8.3354-i2.5170$	&~$3.0535-i0.1510$\\
      CMR~\cite{Guo_CMR}              &$0.3865-i0.1674$	&~$2.2776-i3.7369$  &~$1.3181-i2.7176$ &~$2.9665-i0.5211$	&~$8.3354-i2.5169$	&~$3.0534-i0.1510$	 \\	
      ACCC~\cite{PRC2015Xu_92_024324} &$0.4616-i0.0954$	&~$2.3012-i2.5625$  &~$1.6868-i1.5639$ &~$2.5477-i0.9882$	&~$8.2797-i2.6452$	&~$3.1054-i0.1479$\\
      SPS~\cite{PRC2010LiZP_81_034311}&$0.3750-i0.1719$	&~$-$               &~$-$              &~$2.9500-i0.5362$	&~$8.2200-i2.9806$	&~$3.1000-i0.1579$\\	     \hline\hline
\end{tabular}
\end{table*}

In Table~\ref{Tab2}, we list the energies $E-i\Gamma/2$ of the single-neutron resonant states obtained by solving the coupled-channel Dirac equation using the GF method, in comparison with the results calculated by the CSM~\cite{PRC2017Fang_95_024311}, the ACCC~\cite{PRC2015Xu_92_024324}, and the SPS~\cite{PRC2010LiZP_81_034311} method. The deformation parameter $\beta=0.47$ is chosen in all the calculations and the number of coupled partial waves $N=8$ is taken in the GF, ACCC and SPS calculations. In Table~\ref{Tab2}, it can be seen that obvious differences  exist in the results of the GF, ACCC, and SPS methods although all of them solve the coupled channel Dirac equation in the coordinate space.
Conversely, despite that GF and CSM solve the Dirac equation totally differently, exactly the same energies and widths are obtained for most of the resonant states.
The very slight differences of energies obtained by GF and CSM methods for the broad resonate states may be caused by the different numerical details of the two methods. According to the results in Table~\ref{Tab2}, the resonant states predicted by the GF method in a quadrupole-deformed coupled-channel Dirac equation are also reliable.

\begin{figure}[!]
\includegraphics[width=\linewidth]{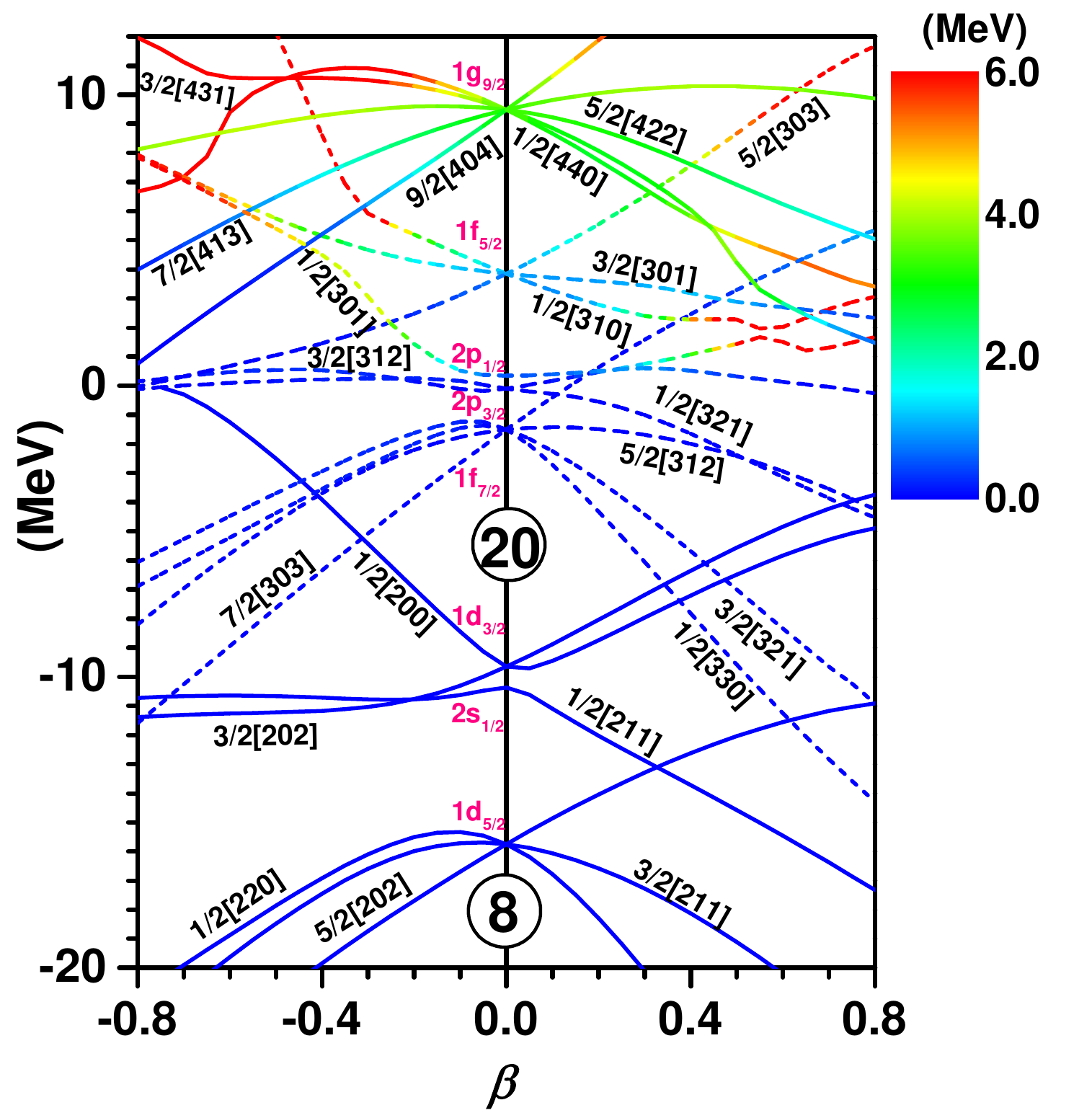}
\caption{\label{fig5}(Color online) Single-neutron levels $\Omega[\mathcal{N} n_{z} \Lambda]$ as a function of deformation parameter $\beta$
in quadrupole-deformed Woods-Saxon potentials. The widths of states $\Gamma$ are represented with different colors.}
\end{figure}

Finally, we apply the deformed GF-Dirac model to investigate the recently reported halo candidate nucleus $^{37}$Mg~\cite{PRL2014Kobayashi_112_242501}, in which the interplay between deformation and orbital structure near threshold is very important. In Fig.~\ref{fig5}, the single-neutron levels for bound and resonant obitals are presented as a function of the deformation parameter $\beta$ in quadrupole-deformed Woods-Saxon potentials. Different colors indicate the widths of the levels in the region $\Gamma\leq6~$MeV. In the spherical case with $\beta=0$, obvious energy gaps $N=8$ and $20$ are obtained while the energy gap $N=28$ disappears as the energy difference between the states $1f_{7/2}$ and $2p_{3/2}$ is only $\sim1.5$~MeV. This is consistent with the results obtained by the ACCC calculations in Ref.~\cite{PRC2015Xu_92_024324} and the systematic calculations on $N=28 $ isotones by the triaxial relativistic Hartree-Bogoliubov model with DD-PC1 density functional~\cite{PRC2011Li_84_054304}. To study the halo phenomena, the weakly bound states and low-lying narrow resonant states near the threshold are of great importance, such as the levels $1/2[321]$, $3/2[312]$, and $1/2[301]$ in Fig.~\ref{fig5}. In particular, a crossing phenomenon between the configurations $1/2[321]$ and $5/2[312]$ happens at a deformation of $\beta\approx 0.5$, which may enhance the probability to occupy the $1/2[321]$ orbital coming from the $2p_{3/2}$ shell and explain the recent observation of a $p$-wave one-neutron halo configuration in $^{37}$Mg~\cite{PRL2014Kobayashi_112_242501}. Similar conclusions are obtained in the studies using the ACCC method~\cite{PRC2015Xu_92_024324} and the coupled-channel Schr\"{o}dinger equation by Hamamoto~\cite{PRC2007Hamamoto_76_054319}, though some differences still exist. The GF function method can describe the single-particle levels very well, which help us analyze the halo structure.

\section{\label{sec:Sum}Summary and Perspectives}


In this work, we apply the Green's function method to investigate the resonant states in quadrupole-deformed Woods-Saxon potentials by solving coupled-channel Dirac equations for the first time. The detailed formalism for the construction of the Green's function on the basis of the coupled-channel representation is presented in coordinate space. Besides, a direct and effective approach is proposed to find resonance parameters of all kinds by searching for the extremes of the density of states, which exactly correspond to the resonant states.

To verify the performance of the deformed GF-Dirac code, numerical checks are carried out in two steps. Firstly, we solve the coupled-channel Dirac equation with the deformation parameter $\beta=0$ and compare the obtained density of states for $\varepsilon > 0$ to those by solving the spherical radial Dirac equation. In addition the energies and widths of resonant states are compared with those obtained by the CSM approach. The excellent agreement between the  results obtained with the two methods demonstrates that the deformed GF-Dirac model is fully correct in the decoupled case. Secondly, for a specific deformed potential with $\beta=0.47$, we carry out the calculations using the GF method and compare the results with those obtained by the CSM, ACCC, and SPS methods. The good agreement between GF and CSM indicates that the GF method is effective for predicting both narrow and wide resonant states in a quadrupole-deformed potential.

Finally, as an application, we investigate the recently reported halo candidate nucleus $^{37}$Mg~\cite{PRL2014Kobayashi_112_242501} and present the evolution of the single-neutron levels for bound and resonant states as a function of the deformation parameter. The pattern of the Nilsson levels calculated by the deformed GF-Dirac model is consistent with those by the deformed ACCC method~\cite{PRC2015Xu_92_024324} and the coupled-channel Schr\"{o}dinger equation~\cite{PRC2007Hamamoto_76_054319}. It is found that a crossing phenomenon between the configurations $1/2[321]$ and $5/2[312]$ happens at a deformation of approximately $0.5$, which may enhance the probability to occupy the $1/2[321]$ orbital coming from $2p_{3/2}$ and explain the observation of a $p$-wave one-neutron halo configuration in $^{37}$Mg.

This work is the first step in an application of the GF method in deformed CDFT. The GF method has been approved to be very effective for the description of resonant states in the continuum within the coupled channel scheme. It is emphasized that the GF method can also describe the wave functions very well by treating the bound and resonant states on the same footing. In other words, the method is very convenient to be used in combination with nuclear models. In the future, we will go further and apply the GF method to the deformed relativistic-Hartree-Bogoliubov theory to describe deformed halos and to the deformed quasiparticle random-phase approximation theory to describe nuclear excitation modes.

\begin{acknowledgements}
Helpful discussions with Professor J.~Y. Guo and S.-H. Ren are appreciated. This work was partly supported by the National Natural Science Foundation of China (Grant Nos.~11875225, 11505157, and 11475140), the Physics Research and Development Program of Zhengzhou University (Grant No.~32410217), and by the DFG cluster of excellence \textquotedblleft
Origins\textquotedblright\ (www.origins-cluster.de).
\end{acknowledgements}

\end{document}